\documentclass[aps,preprintnumbers,amsmath,amssymb,superscriptaddress,floatfix,nofootinbib]{revtex4}
\usepackage{lipsum}
\usepackage{cancel}
\usepackage{slashed}
\usepackage{graphicx}
\usepackage{epsfig}
\usepackage{epstopdf}
\usepackage{hyperref}
\usepackage{amsmath}
\usepackage{amsfonts}
\usepackage{amssymb}
\usepackage{setspace}
\usepackage{amsfonts,color}
\usepackage{caption}
\usepackage{natbib}
\usepackage{subfig}
\usepackage{color}
\usepackage{bbold}
\usepackage{bm}
\usepackage{mathrsfs}
\usepackage{url}
\usepackage{rotating}
\usepackage{pdflscape}
\usepackage{diagbox}
\usepackage[normalem]{ulem}

\usepackage{multirow}
\usepackage{hhline}

\begin{document}


\title{Odderon as Regge oddball spin-3 in $pp$ and $p\bar p$ elastic scattering}

\author{Jingle B. Magallanes}
\email{jingle.magallanes@g.msuiit.edu.ph}
\affiliation{Department of Physics, Mindanao State University - Iligan Institute of Technology, Iligan City, 9200, Philippines}
\author{Prin Sawasdipol}
\email{p.namwongsa@kkumail.com}
\affiliation{Khon Kaen Particle Physics and Cosmology Theory Group (KKPaCT), Department of Physics, Faculty of Science, Khon Kaen University,123 Mitraphap Rd., Khon Kaen, 40002, Thailand}
\author{Chakrit Pongkitivanichkul}
\email{chakpo@kku.ac.th}
\affiliation{Khon Kaen Particle Physics and Cosmology Theory Group (KKPaCT), Department of Physics, Faculty of Science, Khon Kaen University,123 Mitraphap Rd., Khon Kaen, 40002, Thailand}
\author{Daris Samart}
\email{darisa@kku.ac.th, corresponding author}
\affiliation{Khon Kaen Particle Physics and Cosmology Theory Group (KKPaCT), Department of Physics, Faculty of Science, Khon Kaen University,123 Mitraphap Rd., Khon Kaen, 40002, Thailand}

\date{\today}

\begin{abstract}
In this work, we propose that the odderon is a Regge odd-glueball tensor spin-3. To demonstrate our proposal, we study the $pp$ and $p\bar p$ elastic scattering by including contributions of the spin-3 odderon and spin-2 pomeron exchange in the processes. The phenomenological effective Lagrangian approach is used to calculate the $pp$ and $p\bar p$ elastic scattering amplitudes at the tree level. In addition, a Donnachie-Landschoff ansatz of the odderon and pomeron propagators has been used in this work. We fit the theoretical results with the various experimental data of the $pp$ and $p\bar p$ scattering at the TeV scale to determine the model parameters in the present work. By using the model parameters, the Chew-Frautschi plot of the tensor odderon Regge trajectory is evaluated. As a result, the odderon spin-3 mass is predicted to be 3.2 GeV. In addition, the total cross-section of our model is compatible with the results from TOTEM and its extrapolation from D0 collaboration. Moreover, the total cross-section also satisfies the Friossart bound at the Regge limit.   
\end{abstract}

\maketitle
\section{Introduction}\label{sec-1}
Quantum ChromoDynamics (QCD) is a modern theory of strong interaction based on non-abalien color SU(3) quantum gauge field theory describing the interaction of quarks and gluons. QCD is highly successful in explaining hadronic structures and interactions at high energies (momentum exchange) where the strong coupling is small and perturbative quantum field theory is applied. However, at a low energy regime, QCD is a strongly coupled theory that we can not use the standard perturbative theory. 
On the other hand, hadron-hadron scattering in high center of mass energy ($\sqrt{s}$) but low momentum exchange ($t$) known as soft-high energy regime or a Regge limit of $s\to\infty$ and $s\ggg t$, the perturbative QCD is also inapplicable. Before the birth of QCD, Regge theory is invented to describe the hadron-hadron collisions by using analytical properties of the scattering amplitudes \cite{Eden:1966dnq} including a consideration of the complex angular momentum \cite{Gribov:2003nw}. In the Regge theory, the amplitudes of the hadronic processes are scaled as $s^J$ where $J$ is the spin of the exchange particles called Reggeons with fixed relevant quantum numbers \cite{Collins:1977jy}. One can write down the spin as a linear function of $t$ as $J=\alpha(t)$ in the complex angular momentum plane and the linear function $\alpha(t) = \alpha_0 + \alpha'\,t$ is called a Regge trajectory. In addition, the poles (Regge poles) correspond to the families of the exchange particles with increasing spins along the trajectories. As a result, the amplitudes of Regge theory are represented in terms of a sum over all possible exchange particles lying on the Regge trajectory. The cross-sections of various hadronic processes in the soft-high energy scattering limit are successfully described by the Regge theory \cite{Eden:1971jm,Collins:1971ff,Chiu:1972xu,Irving:1977ea}.

According to the experimental data of hadron-hadron scattering in the Regge limit, the total cross-sections slowly grew up with the increase of $s$ whereas the Regge trajectories of all known mesons are not sufficient to explain the experimental data. Then, a so-called pomeron was introduced to address this problem \cite{Chew:1961ev,Gribov:1961ex}. The pomeron is a Reggeon carrying all even charge transformations, vacuum quantum number, with the intercept of Regge trajectory $\alpha_0\approx 1$. This yields the slow growth of the total cross-section at large $s$ \cite{Donnachie:2002en,Forshaw:1997dc}. Various approaches are trying to extract the information of the pomeron trajectory. 
{The typical values of the parameters of the pomeron trajectory from \cite{Collins:1974en,Donnachie:1984xq,Donnachie:1992ny} are $\alpha_0 \approx 1.06-1.08$ and $\alpha'\approx 0.025$ GeV$^{-2}$ \cite{Collins:1974en,Donnachie:1984xq,Donnachie:1992ny}.} The pomeron is generally considered as a bound state of the gluons (glueball). On the other hand, the odd charge-conjugation counterpart of the pomeron called odderon has been proposed by Ref.\cite{Lukaszuk:1973nt}. Similar to the pomeron, the odderon is considered as the glueball with the odd number of gluon compositions. The odderon might cause the different observables between $pp$ and $p\bar p$ due to its charge-conjugation property which is compatible with the experiment. However, the nature and properties of the pomeron and odderon are still unclear so far. A number of approaches have been used to calculate the properties (mass, spin, Regge trajectory and etc.) of pomeron and odderon as glueballs \cite{Akkelin:1990vt,Kuraev:1977fs,Balitsky:1978ic,Kuraev:1976ge,Lipatov:1990zb,Narison:1984hu,Narison:1996fm,Tang:2015twt,Meyer:2004jc,Morningstar:1999rf,Ochs:2013gi,Cotanch:2006wv,Kaidalov:2005kz,Kaidalov:1999de,Kaidalov:1999yd,Llanes-Estrada:2000ozq,Brower:2006ea,Brower:2008cy,Boschi-Filho:2002wdj,Boschi-Filho:2005xct,Colangelo:2007pt,Li:2013oda,Brower:2014wha,Ballon-Bayona:2015wra,FolcoCapossoli:2015jnm,FolcoCapossoli:2016fzj,Dymarsky:2022ecr,Meyer:2004gx,Gregory:2012hu,Chen:2005mg,Llanes-Estrada:2005bii,Mathieu:2008me}. A study of pomeron and odderon exchanges in $pp$ and $p\bar p$ elastic scatterings in the soft-high energy regime has been extensively investigated in various frameworks for instances, phenomenological approaches \cite{Ewerz:2013kda,Ewerz:2016onn, Covolan:1996uy,Block:2012nj,Szanyi:2019kkn,Jenkovszky:2020jca,Broniowski:2018xbg,Csorgo:2020wmw,Csorgo:2019ewn,Csorgo:2018uyp,Ster:2015esa,Block:1984ru,Khoze:2018bus}, QCD inspired models \cite{Halzen:1992vd,Donnachie:1983ff,Ma:2001ji,Hu:2002zr,He:2003fq,Hu:2008zze,Zhou:2006zx,Lu:2020wpo,Bartels:1999yt}, holographic QCD or AdS/CFT correspondence \cite{Domokos:2009hm,Domokos:2010ma,Avsar:2009hc,Hu:2017iix,Xie:2019soz,Burikham:2019zbo,Liu:2022zsa,Liu:2022out,Ballon-Bayona:2017vlm,Iatrakis:2016rvj}. 
 
Recently, however, TOTEM and D0 collaborations have confirmed the existence of the odderon by comparing the experimental data between $pp$ (extrapolated from previous several data) and $p\bar p$ at 1.96 TeV \cite{D0:2012erd,TOTEM:2020zzr}. This reveals the contributions of the odderon in $t$-channel elastic scattering. After the TOTEM and D0 collaborations claimed the discovery of the odderon, several works have done to investigate the properties and scattering processes of the odderon \cite{Chen:2021bck,Chen:2021cjr,Zhang:2021itx,Bence:2021uji,Capossoli:2021ope,Lebiedowicz:2021cgm,Baldenegro:2022xrj,Cui:2022dcm,Lebiedowicz:2022nnn,Bonanno:2022vot}. 
 
Based on the discovery of the odderon and the relevant literature on the field theoretical framework in Ref. \cite{Ewerz:2013kda}, we propose the odderon as a spin-3 tensor odd-glueball within the standard field theoretical framework. We study its consequences in $pp$ and $p\bar p$ elastic scatterings. According to a constituent quark model, the odderon is composed of three-gluon, and the lightest trajectory of the odderon is the spin-3, not the spin-1. This is because the odderon begins with a $J^{PC}$ $= 3^{--}$ three-gluon $L=0$ state with a maximum spin of 3. Similarly, the two-gluon bound state or pomeron starts with the s-wave, and the spin and $PC$ quantum number are assigned as $J^{PC}= 2^{++}$ glueball. Furthermore, a combined lattice QCD calculation and field theoretical Coulomb gauge QCD model confirmed that the odderon can be the oddball starting its Regge trajectory with $J^{PC}$ $= 3^{--}$ \cite{Llanes-Estrada:2005bii}. Various theoretical approaches have also shown that the pomeron is likely to be the spin-2 tensor glueball instead of the scalar one \cite{Domokos:2009hm,Ewerz:2013kda,Ewerz:2016onn,Szanyi:2019kkn,Jenkovszky:2020jca,Lu:2020wpo,Ma:2001ji,Hu:2017iix,Xie:2019soz,Burikham:2019zbo,Liu:2022zsa}. Then the lowest tensor odderon in its Regge trajectory is spin-3 as explained previously. However, the slope, intercept, and mass of the odderon are not well understood. 

In this work, we investigate the elastic scattering of $pp$ and $p\bar p$ with the contributions of the odderon spin-3. The effective Lagrangian of the spin-3 odderon with protons is constructed by using a similar approach in Ref.\cite{Ewerz:2013kda}. Especially, a so-called Donnachie-Landschoff ansatz is used to represent the odderon and pomeron propagators. The contributions of the spin-2 pomeron exchange are also included in the calculation where the effective Lagrangian of pomeron and protons is taken from Ref.\cite{Ewerz:2013kda}. Then the differential cross-sections of the $pp$ and $p\bar p$ elastic scattering are calculated. After a careful statistical analysis, we fit the parameters of our model with several relevant experimental data at the TeV scale. All Feynman rules of our model such as vertices, propagators and etc., can be computed directly from the effective Lagrangians in the conventional method of perturbative QFT. 
The aim of the present work is to make a clear and systematic calculation in order to obtain the amplitudes. The analysis is made with the intention of compatibility with other field theoretical models.
 
 The present work is organized as follows, in the section \ref{sec-2}, we set up our model for $pp$ and $p\bar p$ scattering with pomeron spin-2 and odderon spin-3 exchanges. The amplitudes are also computed as well. The observables will be calculated and free parameters of our model will be fitted with the relevant experimental data at the TeV scales in the section \ref{sec-3}. In section \ref{sec-4}, we close this work by giving discussions and conclusions.

\section{Formalisms: Model set up and scattering amplitudes}\label{sec-2}

\subsection{Effective Lagrangians of the $pp$ and $p\bar p$ scattering in spin-2 pomeron and spin-3 odderon exchange picture}
In this section, we will set up the effective Lagrangians of the $pp$ and $p\bar p$ scattering. It is well-known that the pomeron exchange plays a major role in elastic proton-proton scattering at high energy but low momentum exchange regimes. This work we assume the pomeron as spin-2 tensor particle and the Lagrangian is given by \cite{Ewerz:2013kda,Ewerz:2016onn}
\begin{eqnarray}
\mathscr{L}_{\mathbb{P}} = -i\,g_{\mathbb{P}}\,\mathbb{P}^{\mu\nu}\,\mathcal{G}_{\mu\nu}^{\mu'\nu'}\bar\psi\gamma_{\mu'}\,\overset{\leftrightarrow}{\partial}_{\nu'}\psi \,,
\label{pomeron-Lagrangian}
\end{eqnarray}
where $\mathbb{P}^{\mu\nu} $ is symmetric spin-2 tensor field, $\psi$ is Dirac proton field,  $\bar\psi\,\overset{\leftrightarrow}{\partial}_\mu\,\psi \equiv (\partial_\mu\bar\psi)\,\psi - \bar\psi\,\partial_\mu\psi$\, and the coupling $g_{\mathbb{P}} = 3\times 1.87$ GeV$^{-1}$ as used in Ref.\cite{Ewerz:2013kda}. The totally symmetric tensor $\mathcal{G}_{\mu\nu}^{\mu'\nu'}$ is defined by 
\begin{eqnarray}
\mathcal{G}_{\mu\nu}^{\mu'\nu'} 
= \frac{1}{2!} \left( g_{\mu}^{\mu'}g_{\nu}^{\nu'} + g_{\nu}^{\mu'}g_{\mu}^{\nu'} \right) \,.
\end{eqnarray}
The corresponding vertex function of the pomeron-proton-proton coupling is given by
\begin{eqnarray}
i\Gamma_{\mu\nu}^{\mathbb{P}}(q,q') = -i\,g_{\mathbb{P}}\,\mathcal{G}_{\mu\nu}^{\mu'\nu'}\gamma_{\mu'}\,\big(q'_{\nu'}+ q_{\nu'}\big),
\label{pomeron-vertex}
\end{eqnarray}
where $q$ and $q'$ are incoming and outgoing proton/anti-proton momenta respectively. 

For the spin-3 odderon exchange interaction, the effective Lagrangian reads,
\begin{eqnarray}
\mathscr{L}_{\mathbb{O}} &=& -\,\frac{g_{\mathbb{O}}}{\widetilde{M}_0^{}}\,\mathbb{O}^{\mu\nu\rho}\,\mathcal{G}_{\mu\nu\rho}^{\mu'\nu'\rho'}\,\bar \psi\,\gamma_{\mu'}\,\overset{\leftrightarrow}{\partial}_{\nu'}\,\overset{\leftrightarrow}{\partial}_{\rho'}\,\psi\,,
\label{vector-model}
\end{eqnarray}
where $\mathbb{O}^{\mu\nu\rho}$ is the spin-3 tensor field that totally symmetric under interchanging of the Lorentz indices, i.e., $\mathbb{O}^{\mu\nu\rho}= \mathbb{O}^{\nu\rho\mu} = \mathbb{O}^{\rho\mu\nu} = \mathbb{O}^{\mu\rho\nu} =\mathbb{O}^{\nu\mu\rho} = \mathbb{O}^{\rho\nu\mu}$. In order to obtain the effective Lagrangian in Eq. (\ref{vector-model}), in addition, we have followed the construction of the higher spin field coupling to the nucleons in Ref.\cite{Ewerz:2013kda} by adding the twist-2 operator as shown in Appendix B of \cite{Ewerz:2013kda} and all detail discussions therein. Moreover, the coupling $g_{\mathbb{O}}$ is 
a free parameter in this work and it carries the mass dimension the same as introduced for $g_{\mathbb{P}}$. The ${\widetilde{M}_0^{}}$ is the mass parameter with ${\widetilde{M}_0^{}} = 1$ GeV. In the latter, we will see that this parameter is absent in the scattering amplitude and it is introduced for proper mass dimension. While the totally symmetric tensor $\mathcal{G}_{\mu\nu\rho}^{\mu'\nu'\rho'}$ is used to ensure that the lower indices $(\mu,\nu,\rho)$ of the vertex functions, $\Gamma^{\mathbb{O}}_{\mu\nu\rho}$ is totally symmetric and it is defined by
\begin{eqnarray}
\mathcal{G}_{\mu\nu\rho}^{\mu'\nu'\rho'} = \frac{1}{3!}\left( g_{\mu}^{\mu'}\,g_{\nu}^{\nu'}\,g_{\rho}^{\rho'} + g_{\nu}^{\mu'}\,g_{\rho}^{\nu'}\,g_{\mu}^{\rho'} + g_{\rho}^{\mu'}\,g_{\mu}^{\nu'}\,g_{\nu}^{\rho'} + g_{\nu}^{\mu'}\,g_{\mu}^{\nu'}\,g_{\rho}^{\rho'} + g_{\mu}^{\mu'}\,g_{\rho}^{\nu'}\,g_{\nu}^{\rho'} + g_{\rho}^{\mu'}\,g_{\nu}^{\nu'}\,g_{\mu}^{\rho'}\,. \right)
\end{eqnarray}
According to the Lagrangian in Eqs.(\ref{vector-model}), we can write the Feynman rules for the vertex function for $\mathscr{L}_{\mathbb{O}}$ as
\begin{eqnarray}
i\Gamma^{\mathbb{O}}_{\mu\nu\rho}(q,q') &=& -i \frac{g_{\mathbb{O}}}{\widetilde{M}_0^{}}\,\mathcal{G}_{\mu\nu\rho}^{\mu'\nu'\rho'}\,\gamma_{\mu'}\,(q'_{\nu'}+q_{\nu'})\,(q'_{\rho'} + q_{\rho'})\,,
\label{vector-vertex}
\end{eqnarray}
where $q$ and $q'$ represent the incoming and outgoing momenta of the proton/anti-proton of the vertex functions. 
\subsection{Scattering amplitudes of the $pp$ and $p\bar p$ elastic scattering}
Next step, we will calculate the amplitudes for the elastic $pp$ and $p\bar p$ scattering processes under the external momentum specifications as $p(q_1)\,p(q_2)\to p(q_3)\,p(q_4)$ and $\bar p(q_1)\,p(q_2)\to \bar p(q_3)\,p(q_4)$ for $pp$ and $p\bar p$ elastic scattering processes, respectively. In addition, we assume that the elastic $pp$ and $p\bar p$ scatterings are mainly dominated by the pomeron and odderon exchanges since other mesons and Reggeons exchange contributions are negligibly small in these processes at TeV scale.
\\\\
By using the standard method in QFT \cite{Peskin:1995ev}, the elastic $pp$ scattering amplitude of the pomeron exchange is given by
\begin{eqnarray}
i\,\mathcal{M}_{\mathbb{P}}^{pp} &=& \langle p(q_3)\,p(q_4)\,
|:\mathcal{T}\,\exp{\left(i\int \mathscr{L}_{\mathbb{P}}(x)\,d^4x\right)}:|\,p(q_1)\,p(q_2)\rangle
\nonumber\\
&=& -i\,g_{\mathbb{P}}^2\,\mathcal{G}_{\mu_1\nu_1}^{\mu_1'\nu_1'}\,\bar u(q_3)\,\gamma_{\mu_1'}\,\big(q_{3,\nu_1'}+ q_{1,\nu_1'}\big)\,u(q_1)\,i\Delta_{\mathbb{P}}^{\mu_1\nu_1;\mu_2\nu_2}(s,t)
\nonumber\\
&&\qquad\times\,\mathcal{G}_{\mu_2\nu_2}^{\mu_2'\nu_2'}\,\bar u(q_4)\,\gamma_{\mu_2'}\,\big(q_{4,\nu_2'}+ q_{2,\nu_2'}\big) \,u(q_2)\,\mathcal{F}_{\mathbb{P}}(t)^2\,,
\label{amp-pomeron-pp}
\end{eqnarray}
and the elastic $p\bar p$ scattering amplitude of the pomeron exchange is given by
\begin{eqnarray}
i\,\mathcal{M}_{\mathbb{P}}^{p\bar p} &=& \langle \bar p(q_3)\,p(q_4)\,
|:\mathcal{T}\,\exp{\left(i\int \mathscr{L}_{\mathbb{P}}(x)\,d^4x\right)}:|\,\bar p(q_1)\,p(q_2)\rangle
\nonumber\\
&=& -i\,g_{\mathbb{P}}^2\,\mathcal{G}_{\mu_1\nu_1}^{\mu_1'\nu_1'}\,\bar v(q_1)\,\gamma_{\mu_1'}\,\big(q_{1,\nu_1'}+ q_{3,\nu_1'}\big)  \,v(q_3)\,i\Delta_{\mathbb{P}}^{\mu_1\nu_1;\mu_2\nu_2}(s,t)
\nonumber\\
&&\qquad\times\,\mathcal{G}_{\mu_2\nu_2}^{\mu_2'\nu_2'}\,\bar u(q_4)\,\gamma_{\mu_2'}\,\big(q_{4,\nu_2'}+ q_{2,\nu_2'}\big) \,u(q_2)\,\mathcal{F}_{\mathbb{P}}(t)^2\,.
\label{amp-pomeron-pbarp}
\end{eqnarray}
For the propagator of the spin-2 pomeron $\Delta_{\mathbb{P}}^{\mu\nu;\mu'\nu'}$, we employ from Ref.\cite{Ewerz:2013kda} and it takes the following form 
\begin{eqnarray}
i\Delta^{\mu\nu,\rho\sigma}_{\mathbb{P}}(s,t) &=& \frac{1}{4\,s}\Big[ g^{\mu\rho}\,g^{\nu\sigma} + g^{\mu\sigma}\,g^{\nu\rho} -\frac12\,g^{\mu\nu}\,g^{\rho\sigma}\Big]\,\left( -i\,\alpha_{\mathbb{P}}'\,s\right)^{\alpha_{\mathbb{P}}(t)-1}\,,
\label{pom-prop}
\end{eqnarray}
with the conventional linear pomeron trajectory \cite{Ewerz:2013kda,Ewerz:2016onn}
\begin{eqnarray}
\alpha_{\mathbb{P}}(t) &=& 1 + \epsilon_{\mathbb{P}} + \alpha_{\mathbb{P}}'\,t\,,
\\
\epsilon_{\mathbb{P}} &=& 0.0808\,, \qquad {\rm and}\qquad \alpha_{\mathbb{P}}'= 0.25 \;{\rm GeV}^{-2},
\end{eqnarray}
where the factor $1+\epsilon_{\mathbb{P}}$ and $\alpha_{\mathbb{P}}'$ represent the vertical interception and slope of the pomeron trajectory, respectively.
This formulation of the spin-2 pomeron propagator is an alternative approach to studying soft high energy hadronic collisions. This formalism can be reproduced in several experimental data. In addition, the pomeron propagator with the Donnachie-Landschoff ansatz in Eq.(\ref{pom-prop}) is proposed by Heidelberg group \cite{Ewerz:2013kda}. Finally, the pomeron-$pp/p\bar p$ coupling form factor reads,
\begin{eqnarray}
\mathcal{F}_{\mathbb{P}}(t) &=& \left(1-\frac{t}{4m_p^2}\frac{\mu_p}{\mu_N}\right)\left( 1 - \frac{t}{4m_p^2}\right)^{-1}\left( 1 - \frac{t}{m_D^2}\right)^{-2} ,
\label{pomeron-FF}
\nonumber\\
\mu_N &=& \frac{e}{2m_p}\,,\qquad \frac{\mu_p}{\mu_N} = 2.7928\,, \qquad m_D^2 = 0.71\;{\rm GeV}^2\,.
\end{eqnarray}
This form factor is the standard Dirac form factor of proton and it is widely used to study the $pp$ and $p\bar p$ scattering see more details discussions and its consequences of the form factor in Eq.(\ref{pomeron-FF}) in chapter two of Ref.\cite{Close:2007zzd}.  
\\
\\
We turn to consider the elastic $pp$ and $p\bar p$ scattering amplitudes for the spin-3 odderon exchange contribution. Having used the same manner, the $pp$ scattering amplitude is calculated and one finds,
\allowdisplaybreaks
\begin{eqnarray}
i\mathcal{M}_{\mathbb{O}}^{pp} &=& \langle p(q_3)\,p(q_4)\,
|:\mathcal{T}\,\exp{\left(i\int \mathscr{L}_{\mathbb{O}}(x)\,d^4x\right)}:|\,p(q_1)\,p(q_2)\rangle
\nonumber\\
&=& -\,\frac{g_{\mathbb{O}}^2}{\widetilde{M}_0^2}\,\mathcal{G}_{\mu\nu\rho}^{\mu_1\nu_1\rho_1}\,\bar u(q_3)\,\gamma_{\mu_1}\,\big(q_{3,\nu_1} + q_{1,\nu_1}\big)\,\big(q_{3,\rho_1} + q_{1,\rho_1}\big)\, u(q_1)\,i\Delta_{\mathbb{O}}^{\mu\nu\rho;\mu'\nu'\rho'}\big(s,t\big)\,
\nonumber\\
&\;&\quad\;\;\, \times\,\mathcal{G}_{\mu'\nu'\rho'}^{\mu_2\nu_2\rho_2}\,\bar u(q_4)\,\gamma_{\mu_2}\,\big(q_{4,\nu_2} + q_{2,\nu_2}\big)\,\big(q_{4,\rho_2} + q_{2,\rho_2}\big)\,u(q_2)\,\mathcal{F}_{\mathbb{O}}(t)^2\,,
\label{amp-vector-pp}
\end{eqnarray}
while the amplitude of the $p\bar p$ scattering with the odderon exchange reads
\begin{eqnarray}
i\mathcal{M}_{\mathbb{O}}^{p\bar p} &=& \langle \bar p(q_3)\,p(q_4)\,
|:\mathcal{T}\,\exp{\left(i\int \mathscr{L}_{\mathbb{O}}(x)\,d^4x\right)}:|\,\bar p(q_1)\,p(q_2)\rangle
\nonumber\\
&=& -\,\frac{g_{\mathbb{O}}^2}{\widetilde{M}_0^2}\,\mathcal{G}_{\mu\nu\rho}^{\mu_1\nu_1\rho_1}\,\bar v(q_1)\,\gamma_{\mu_1}\,\big(q_{1,\nu_1} + q_{3,\nu_1}\big)\,\big(q_{1,\rho_1} + q_{3,\rho_1}\big)\, v(q_3)\,i\Delta_{\mathbb{O}}^{\mu\nu\rho;\mu'\nu'\rho'}\big(s,t\big)
\nonumber\\
&\;&\quad\;\;\, \times\,\mathcal{G}_{\mu'\nu'\rho'}^{\mu_2\nu_2\rho_2}\,\bar u(q_4)\,\gamma_{\mu_2}\,\big(q_{4,\nu_2} + q_{2,\nu_2}\big)\,\big(q_{4,\rho_2} + q_{2,\rho_2}\big)\,u(q_2)\,\mathcal{F}_{\mathbb{O}}(t)^2\,.
\label{amp-vector-pbarp}
\end{eqnarray}
In addition, we have modified the proton form-factor in Eq.(\ref{pomeron-FF}) for the odderon coupling to $pp$ by adding new three free parameters, $A$, $B$ and $C$ as,
\begin{eqnarray}
\mathcal{F}_{\mathbb{O}}(t) &=& \left(1-\frac{A\,t}{4m_p^2}\frac{\mu_p}{\mu_N}\right)\left( 1 - \frac{B\,t}{4m_p^2}\right)^{-1}\left( 1 - \frac{C\,t}{m_D^2}\right)^{-2} .
\end{eqnarray} 
The $\Delta_{\mathbb{O}}^{\mu_1\nu_1\rho_1;\mu_2\nu_2\rho_2}(s,t)$ term is the propagator of the spin-3 particle in momentum space. By analogy to the spin-2 tensor pomeron proposed by \cite{Ewerz:2013kda}, we introduce the spin-3 tensor odderon propagator with the Donnachie-Lanschoff parametrization and it reads,
\begin{eqnarray}
i\Delta^{\mu\nu\lambda,\rho\sigma\tau}_{\mathbb{O}}(s,t) &=& {\color{black}-i}\frac{\widetilde{M}_0^2}{6\,s^2}\Big[ \sum_{C}\,g^{\mu\rho}\left(g^{\nu\sigma}\,g^{\lambda\tau} + g^{\nu\tau}\,g^{\lambda\sigma} \right) -\frac12\sum_{C}\,g^{\mu\nu}\,g^{\lambda\rho}\,g^{\sigma\tau}\Big]\,\left( -i\,\alpha_{\mathbb{O}}'\,s\right)^{\alpha_{\mathbb{O}}(t)-1}\,,
\\
\alpha_{\mathbb{O}}(t) &=& 1 + \epsilon_{\mathbb{O}} + \alpha_{\mathbb{O}}'\,t\,,
\end{eqnarray}
where $\sum_C$ stands for the sum over all distinct combinations of the Lorentz indices ($\mu\nu\lambda$) and ($\rho\sigma\tau$), for instance,
\begin{eqnarray}
\sum_{C}\,g_{\mu\nu}\,g_{\lambda\rho}\,g_{\sigma\tau} 
&=& 
g_{\mu\nu}\,g_{\lambda\rho}\,g_{\sigma\tau} + g_{\mu\nu}\,g_{\lambda\sigma}\,g_{\rho\tau} + g_{\mu\nu}\,g_{\lambda\tau}\,g_{\rho\sigma} 
+ 
g_{\mu\lambda}\,g_{\nu\rho}\,g_{\sigma\tau} + g_{\mu\lambda}\,g_{\nu\sigma}\,g_{\rho\tau} + g_{\mu\lambda}\,g_{\nu\tau}\,g_{\rho\sigma} 
\nonumber\\
&+&
g_{\nu\lambda}\,g_{\mu\rho}\,g_{\sigma\tau} + g_{\nu\lambda}\,g_{\mu\sigma}\,g_{\rho\tau} + g_{\nu\lambda}\,g_{\mu\tau}\,g_{\rho\sigma} \,.
\end{eqnarray}
The mass parameter $M_0$ is a free parameter in this work and it is introduced in order to correct the mass dimension of the odderon propagator. In addition, we consider the parameters $\epsilon_{\mathbb{O}}$ and $\alpha_{\mathbb{O}}'$ as free parameters. However, the condition $\epsilon_{\mathbb{O}} < \epsilon_{\mathbb{P}}$ is imposed due to the experimental data fact that the total cross-sections of both $pp$ and $p\bar p$ are identical at very high energies. In the other words, the  pomeron exchange contributions for elastic $pp$ and $p\bar p$ scattering at the Regge limit always dominate over the odderon ones. Moreover, the tensor structure of the spin-3 odderon propagator has been constructed in Ref. \cite{Berends:1979je} (see more detailed derivations and discussions). We close this section by giving the definitions of the four-momentum conservation, the on-shell mass of the particles, and the Mandelstam variables as
\begin{eqnarray}
&& q_1+ q_2 = q_3 + q_4\,,\quad q_1^2 = q_2^2 = q_3^2 = q_4^2 = m_p^2\,,\quad s + t + u = 4\,m_p^2\,, 
\nonumber\\
&& s = (q_1 + q_2)^2 = (q_3 + q_4)^2 \,,\quad t = (q_3 - q_1)^2 = (q_4 - q_2)^2 \,,
\quad u = (q_4 - q_1)^2 = (q_3 - q_2)^2 .
\label{kin-rules}
\end{eqnarray}
In the next section, we will provide analytical expressions of the differential cross-section for the $pp$ and $p\bar p$ elastic scattering with pomeron and odderon exchanges and fit the model parameters with the experimental data of the $pp$ and $p\bar p$ elastic scatterings at TeV scale. 
\section{ Results and Discussions}\label{sec-3}

\subsection{Differential cross section formulae}
In this subsection, we will provide the analytical formulae of the differential cross section with respect to the $t$ variable and then the model parameters in the present work will be determined by fitting with all available experimental data of the $pp$ and $p\bar p$ elastic scatterings at TeV level. The differential cross-section of the $pp$ and $p\bar p$ scattering is given by \cite{Ewerz:2016onn},
\begin{eqnarray}
\frac{d\sigma^{pp/p\bar p}}{dt} = \frac{1}{16\pi s\left( s - m_p^2\right)}\,\frac14\,\sum_{\rm spin}\,\big|\mathcal{M}^{pp/p\bar p}\big|^2.
\end{eqnarray}

First of all, let us briefly discuss the definitions of the scattering amplitudes of $pp$ and $p\bar p$ in the pomeron and odderon exchange picture. Considering the amplitude $\mathcal{M}^{ab}(s,t)$ of an elastic scattering for the $a+b\to a+b$ process in $s$ channel. While the corresponding elastic scattering by crossing to the $u$-channel as $a+\bar b\to a+\bar b$ with the amplitude $\mathcal{M}^{a\bar b}(u,t)$.
According to the crossing symmetry of the scattering amplitudes, they are symmetric under interchange between the Mandelstam variables, $s$ and $u$ as
\begin{eqnarray}
\mathcal{M}^{a\bar b}(s,t,u) = \mathcal{M}^{ab}(u,t,s)\,.
\end{eqnarray}
Moreover, the amplitude $\mathcal{M}_{\pm}$ are defined from $\mathcal{M}^{ab/a\bar b}$ as
\begin{eqnarray}
\mathcal{M}_{\pm}(s,t) = \frac12\left( \mathcal{M}^{ab}(s,t) \pm \mathcal{M}^{a\bar b}(s,t)\right)\,.
\end{eqnarray}
Interchange $s\to u$, one finds that the amplitude $\mathcal{M}_{+}$ is invariant under the crossing symmetry whereas the amplitude $\mathcal{M}_{-}$ changes the relative sign. We therefore call $\mathcal{M}_{+}$ and $\mathcal{M}_{-}$ as even and odd under the crossing symmetry. As a result, one observes that the $\mathcal{M}_{+}$ and $\mathcal{M}_{-}$ correspond to the even and odd under charge conjugation, respectively. Since the interchange $s\to u$ is equivalent to charge conjugation transformation ($\mathcal{C}$) i.e., changing particle-particle scattering to particle-antiparticle scattering. We therefore identify $\mathcal{M}_{+}$ and $\mathcal{M}_{-}$ as pomeron ($\mathcal{M}_{\mathbb{P}}$ with $\mathcal{C} = +1$) and odderon ($\mathcal{M}_{\mathbb{O}}$ with $\mathcal{C} = -1$) exchange amplitudes, respectively. 

As discussed above, we can define the total amplitude of the elastic $pp$ and $p\bar p$ processes with pomeron and odderon exchange diagrams at the tree level as follow
\begin{eqnarray}
\mathcal{M}^{pp} &=& \mathcal{M}_{\mathbb{P}}^{pp} - \mathcal{M}_{\mathbb{O}}^{pp}\,,
\label{pp-amp}
\\
\mathcal{M}^{p\bar p} &=& \mathcal{M}_{\mathbb{P}}^{p\bar p} + \mathcal{M}_{\mathbb{O}}^{p\bar p}\,.
\label{ppbar-amp}
\end{eqnarray}

The absolute square amplitudes of the $\mathcal{M}^{pp}$ and $\mathcal{M}^{p\bar p}$ with an average sum over unpolarized initial spin states of the incoming particles are given by
\begin{eqnarray}
\sum_{\rm spin}\,\big|\mathcal{M}^{pp}\big|^2 &=& \sum_{\rm spin}\left(\big|\mathcal{M}_{\mathbb{P}}^{pp}\big|^2 - \big| \mathcal{M}_{\mathbb{P}}^{pp}\,\mathcal{M}_{\mathbb{O}}^{pp\,*}\big| - \big| \mathcal{M}_{\mathbb{P}}^{pp\,*}\,\mathcal{M}_{\mathbb{O}}^{pp}\big|  + \big|\mathcal{M}_{\mathbb{O}}^{pp}\big|^2\right),
\\
\sum_{\rm spin}\,\big|\mathcal{M}^{p\bar p}\big|^2 &=& 
\sum_{\rm spin}\left(\big|\mathcal{M}_{\mathbb{P}}^{p\bar p}\big|^2 + \big| \mathcal{M}_{\mathbb{P}}^{p\bar p\,*}\,\mathcal{M}_{\mathbb{O}}^{p\bar p}\big| + \big| \mathcal{M}_{\mathbb{P}}^{p\bar p}\,\mathcal{M}_{\mathbb{O}}^{p\bar p\,*}\big| + \big|\mathcal{M}_{\mathbb{O}}^{p\bar p}\big|^2\right).
\end{eqnarray}
The explicit forms of the absolute square amplitudes with an average sum over initial states can be calculated in the following forms,
\begin{eqnarray}
\sum_{\rm spin}\,\big|\mathcal{M}_{\mathbb{P}}^{pp}\big|^2
&=& {\rm Tr}\,\Big[ \big(q\!\!\!/_3+m_p\big)\,\gamma_{\mu_1'}\,\big(q_{3,\nu_1'}+ q_{1,\nu_1'}\big)\,\big(q\!\!\!/_1+m_p\big)\,\gamma_{\mu_3'}\,\big(q_{3,\nu_3'}+ q_{1,\nu_3'}\big)
\nonumber\\
&&\,\times\,\big(q\!\!\!/_4+m_p\big)\,\gamma_{\mu_2'}\,\big(q_{4,\nu_2'}+ q_{2,\nu_2'}\big) \,\big(q\!\!\!/_2+m_p\big)\,\gamma_{\mu_4'}\,\big(q_{4,\nu_4'}+ q_{2,\nu_4'}\big)\Big]
\nonumber\\
&&\,\times\, \,g_{\mathbb{P}}^4\,\mathcal{G}_{\mu_1\nu_1}^{\mu_1'\nu_1'}\,
\mathcal{G}_{\mu_3\nu_3}^{\mu_3'\nu_3'}\,\mathcal{G}_{\mu_2\nu_2}^{\mu_2'\nu_2'}\,\mathcal{G}_{\mu_4\nu_4}^{\mu_4'\nu_4'}\,\Delta_{\mathbb{P}}^{\mu_1\nu_1;\mu_2\nu_2}(s,t)\,\Delta_{\mathbb{P}}^{\mu_3\nu_3;\mu_4\nu_4}(s,t)\,\mathcal{F}_{\mathbb{P}}(t)^4 
\nonumber\\
&\approx& \frac{16\,g_{\mathbb{P}}^4 \left[1 - \frac{t}{4m_p^2}\frac{\mu_p}{\mu_N}\right]^4}{\left[1 - \frac{t}{m_p^2}\right]^4 \left[1 -\frac{t}{m_D^2}\right]^8 }\, s^2  \left(\alpha'_{\mathbb{P}} s\right)^{2\epsilon_{\mathbb{P}} + 2\alpha'_{\mathbb{P}} t },
\label{sq-amp-p-pp}
\\
\sum_{\rm spin}\,\big|\mathcal{M}_{\mathbb{O}}^{pp}\big|^2 &=&
{\rm Tr}\,\Big[\big(q\!\!\!/_3+m_p\big)\,\gamma_{\mu_1}\,\big(q_{3,\nu_1} + q_{1,\nu_1}\big)\,\big(q_{3,\rho_1} + q_{1,\rho_1}\big)\, \big(q\!\!\!/_1+m_p\big)\,\gamma_{\bar\mu_3}\,\big(q_{3,\bar\nu_3} + q_{1,\bar\nu_3}\big)\,\big(q_{3,\bar\rho_3} + q_{1,\bar\rho_3}\big)
\nonumber\\
&&\,\times\,\big(q\!\!\!/_4+m_p\big)\,\gamma_{\mu_2}\,\big(q_{4,\nu_2} + q_{2,\nu_2}\big)\,\big(q_{4,\rho_2} + q_{2,\rho_2}\big)\,\big(q\!\!\!/_2+m_p\big)\,\gamma_{\bar\mu_4}\,\big(q_{4,\bar\nu_4} + q_{2,\bar\nu_4}\big)\,\big(q_{4,\bar\rho_4} + q_{2,\bar\rho_4}\big)\Big]
\nonumber\\
&&\,\times\,\frac{g_{\mathbb{O}}^4}{\widetilde{M}_0^4}\,\mathcal{G}_{\mu\nu\rho}^{\mu_1\nu_1\rho_1}\,\mathcal{G}_{\bar\mu\bar\nu\bar\rho}^{\bar\mu_3\bar\nu_3\bar\rho_3}\,\mathcal{G}_{\mu'\nu'\rho'}^{\mu_2\nu_2\rho_2}\,\mathcal{G}_{\bar\mu'\bar\nu'\bar\rho'}^{\bar\mu_4\bar\nu_4\bar\rho_4}\,\Delta_{\mathbb{O}}^{\mu\nu\rho;\mu'\nu'\rho'}\big(s,t\big)\,\Delta_{\mathbb{O}}^{\bar\mu\bar\nu\bar\rho;\bar\mu'\bar\nu'\bar\rho'}\big(s,t\big)\,\mathcal{F}_{\mathbb{O}}(t)^4 
\nonumber\\
&\approx& \frac{ 64\, g_{\mathbb{O}}^4 \left[1- \frac{At}{4m_p^2}\frac{\mu_p}{\mu_N}\right]^4}{{9} \left[1 - \frac{B t}{m_p^2}\right]^4 \left[ 1 - \frac{C t}{m_D^2}\right]^8} \, s^2 \left(\alpha'_{\mathbb{O}} s\right)^{2\epsilon_{\mathbb{O}} + 2\alpha'_{\mathbb{O}} t },
\label{sq-amp-o-pp}
\\
\sum_{\rm spin}\,\big|\mathcal{M}_{\mathbb{P}}^{pp}\mathcal{M}_{\mathbb{O}}^{pp\,*}\big|
&=& {\rm Tr}\,\Big[ \big(q\!\!\!/_3+m_p\big)\,\gamma_{\mu_1'}\,\big(q_{3,\nu_1'}+ q_{1,\nu_1'}\big)\,\big(q\!\!\!/_1+m_p\big)\,\gamma_{\bar\mu_3}\,\big(q_{3,\bar\nu_3} + q_{1,\bar\nu_3}\big)\,\big(q_{3,\bar\rho_3} + q_{1,\bar\rho_3}\big)
\nonumber\\
&&\times\,\big(q\!\!\!/_4+m_p\big)\,\gamma_{\mu_2'}\,\big(q_{4,\nu_2'}+ q_{2,\nu_2'}\big) \,\big(q\!\!\!/_2+m_p\big)\,\gamma_{\bar\mu_4}\,\big(q_{4,\bar\nu_4} + q_{2,\bar\nu_4}\big)\,\big(q_{4,\bar\rho_4} + q_{2,\bar\rho_4}\big)\Big]
\nonumber\\
&&\times\, \frac{g_{\mathbb{P}}^2\,g_{\mathbb{O}}^2}{\widetilde{M}_0^2}\,\mathcal{G}_{\mu_1\nu_1}^{\mu_1'\nu_1'}\,
\mathcal{G}_{\bar\mu\bar\nu\bar\rho}^{\bar\mu_3\bar\nu_3\bar\rho_3}\,\mathcal{G}_{\mu_2\nu_2}^{\mu_2'\nu_2'}\,\mathcal{G}_{\bar\mu'\bar\nu'\bar\rho'}^{\bar\mu_4\bar\nu_4\bar\rho_4}\,\Delta_{\mathbb{P}}^{\mu_1\nu_1;\mu_2\nu_2}(s,t)\,\Delta_{\mathbb{O}}^{\bar\mu\bar\nu\bar\rho;\bar\mu'\bar\nu'\bar\rho'}\big(s,t\big)\,\mathcal{F}_{\mathbb{P}}(t)^2\mathcal{F}_{\mathbb{O}}(t)^2
\nonumber\\
&\approx& \frac{32\,g_{\mathbb{P}}^2\,g_{\mathbb{O}}^2 \left[1 - \frac{t}{4m_p^2}\frac{\mu_p}{\mu_N}\right]^2 \left[1- \frac{At}{4m_p^2}\frac{\mu_p}{\mu_N}\right]^2}{{9} \left[1 - \frac{t}{m_p^2}\right]^2\left[1 -\frac{t}{m_D^2}\right]^4 \left[1 - \frac{B t}{m_p^2}\right]^2 \left[ 1 - \frac{C t}{m_D^2}\right]^4}\, s^2   \left({-}i\,\alpha'_{\mathbb{P}} s\right)^{\epsilon_{\mathbb{P}} + \alpha'_{\mathbb{P}} t }\left(i\,\alpha'_{\mathbb{O}} s\right)^{\epsilon_{\mathbb{O}} + \alpha'_{\mathbb{O}} t },
\label{sq-amp-po-pp}
\\
\sum_{\rm spin}\,\big|\mathcal{M}_{\mathbb{P}}^{p\bar p}\big|^2
&=& {\rm Tr}\,\Big[ \big(q\!\!\!/_1-m_p\big)\,\gamma_{\mu_1'}\,\big(q_{3,\nu_1'}+ q_{1,\nu_1'}\big)\,\big(q\!\!\!/_3-m_p\big)\,\gamma_{\mu_3'}\,\big(q_{3,\nu_3'}+ q_{1,\nu_3'}\big)
\nonumber\\
&&\times\,\big(q\!\!\!/_4+m_p\big)\,\gamma_{\mu_2'}\,\big(q_{4,\nu_2'}+ q_{2,\nu_2'}\big) \,\big(q\!\!\!/_2+m_p\big)\,\gamma_{\mu_4'}\,\big(q_{4,\nu_4'}+ q_{2,\nu_4'}\big)\Big]
\nonumber\\
&&\times\, g_{\mathbb{P}}^4\,\mathcal{G}_{\mu_1\nu_1}^{\mu_1'\nu_1'}\,
\mathcal{G}_{\mu_3\nu_3}^{\mu_3'\nu_3'}\,\mathcal{G}_{\mu_2\nu_2}^{\mu_2'\nu_2'}\,\mathcal{G}_{\mu_4\nu_4}^{\mu_4'\nu_4'}\,\Delta_{\mathbb{P}}^{\mu_1\nu_1;\mu_2\nu_2}(s,t)\,\Delta_{\mathbb{P}}^{\mu_3\nu_3;\mu_4\nu_4}(s,t)\,\mathcal{F}_{\mathbb{P}}(t)^4 
\nonumber\\
&\approx& \frac{16\,g_{\mathbb{P}}^4 \left[1 - \frac{t}{4m_p^2}\frac{\mu_p}{\mu_N}\right]^4}{\left[1 - \frac{t}{m_p^2}\right]^4 \left[1 -\frac{t}{m_D^2}\right]^8 }\, s^2  \left(\alpha'_{\mathbb{P}} s\right)^{2\epsilon_{\mathbb{P}} + 2\alpha'_{\mathbb{P}} t },
\label{sq-amp-p-ppbar}
\\
\sum_{\rm spin}\,\big|\mathcal{M}_{\mathbb{O}}^{p\bar p}\big|^2 &=&
{\rm Tr}\,\Big[\big(q\!\!\!/_1-m_p\big)\,\gamma_{\mu_1}\,\big(q_{3,\nu_1} + q_{1,\nu_1}\big)\,\big(q_{3,\rho_1} + q_{1,\rho_1}\big)\, \big(q\!\!\!/_3-m_p\big)\,\gamma_{\bar\mu_3}\,\big(q_{3,\bar\nu_3} + q_{1,\bar\nu_3}\big)\,\big(q_{3,\bar\rho_3} + q_{1,\bar\rho_3}\big)
\nonumber\\
&&\times\,\big(q\!\!\!/_4+m_p\big)\,\gamma_{\mu_2}\,\big(q_{4,\nu_2} + q_{2,\nu_2}\big)\,\big(q_{4,\rho_2} + q_{2,\rho_2}\big)\,\big(q\!\!\!/_2+m_p\big)\,\gamma_{\bar\mu_4}\,\big(q_{4,\bar\nu_4} + q_{2,\bar\nu_4}\big)\,\big(q_{4,\bar\rho_4} + q_{2,\bar\rho_4}\big)\Big]
\nonumber\\
&&\times\,\frac{g_{\mathbb{O}}^4}{\widetilde{M}_0^4}\,\mathcal{G}_{\mu\nu\rho}^{\mu_1\nu_1\rho_1}\,\mathcal{G}_{\bar\mu\bar\nu\bar\rho}^{\bar\mu_3\bar\nu_3\bar\rho_3}\,\mathcal{G}_{\mu'\nu'\rho'}^{\mu_2\nu_2\rho_2}\,\mathcal{G}_{\bar\mu'\bar\nu'\bar\rho'}^{\bar\mu_4\bar\nu_4\bar\rho_4}\,\Delta_{\mathbb{O}}^{\mu\nu\rho;\mu'\nu'\rho'}\big(s,t\big)\,\Delta_{\mathbb{O}}^{\bar\mu\bar\nu\bar\rho;\bar\mu'\bar\nu'\bar\rho'}\big(s,t\big)\,\mathcal{F}_{\mathbb{O}}(t)^4 
\nonumber\\
&\approx& \frac{64\,g_{\mathbb{O}}^4 \,\left[1- \frac{At}{4m_p^2}\frac{\mu_p}{\mu_N}\right]^4}{{9} \left[1 - \frac{B t}{m_p^2}\right]^4 \left[ 1 - \frac{C t}{m_D^2}\right]^8} \, s^2 \left(\alpha'_{\mathbb{O}} s\right)^{2\epsilon_{\mathbb{O}} + 2\alpha'_{\mathbb{O}} t },
\label{sq-amp-o-ppbar}\\
\sum_{\rm spin}\,\big|\mathcal{M}_{\mathbb{P}}^{p\bar p}\mathcal{M}_{\mathbb{O}}^{p\bar p\,*}\big|^2
&=& {\rm Tr}\,\Big[ \big(q\!\!\!/_1-m_p\big)\,\gamma_{\mu_1'}\,\big(q_{3,\nu_1'}+ q_{1,\nu_1'}\big)\,\big(q\!\!\!/_3-m_p\big)\,\gamma_{\bar\mu_3}\,\big(q_{3,\bar\nu_3} + q_{1,\bar\nu_3}\big)\,\big(q_{3,\bar\rho_3} + q_{1,\bar\rho_3}\big)
\nonumber\\
&&\times\,\big(q\!\!\!/_4+m_p\big)\,\gamma_{\mu_2'}\,\big(q_{4,\nu_2'}+ q_{2,\nu_2'}\big) \,\big(q\!\!\!/_2+m_p\big)\,\gamma_{\bar\mu_4}\,\big(q_{4,\bar\nu_4} + q_{2,\bar\nu_4}\big)\,\big(q_{4,\bar\rho_4} + q_{2,\bar\rho_4}\big)\Big]
\nonumber\\
&&\times\,  \frac{g_{\mathbb{P}}^2\,g_{\mathbb{O}}^2}{\widetilde{M}_0^2}\,\mathcal{G}_{\mu_1\nu_1}^{\mu_1'\nu_1'}\,
\mathcal{G}_{\bar\mu\bar\nu\bar\rho}^{\bar\mu_3\bar\nu_3\bar\rho_3}\,\mathcal{G}_{\mu_2\nu_2}^{\mu_2'\nu_2'}\,\mathcal{G}_{\bar\mu'\bar\nu'\bar\rho'}^{\bar\mu_4\bar\nu_4\bar\rho_4}\,\Delta_{\mathbb{P}}^{\mu_1\nu_1;\mu_2\nu_2}(s,t)\,\Delta_{\mathbb{O}}^{\bar\mu\bar\nu\bar\rho;\bar\mu'\bar\nu'\bar\rho'}\big(s,t\big)\,\mathcal{F}_{\mathbb{P}}(t)^2\mathcal{F}_{\mathbb{O}}(t)^2 
\nonumber\\
&\approx& \frac{32\,g_{\mathbb{P}}^2 g_{\mathbb{O}}^2 \left[1 - \frac{t}{4m_p^2}\frac{\mu_p}{\mu_N}\right]^2 \left[1- \frac{At}{4m_p^2}\frac{\mu_p}{\mu_N}\right]^2}{{9} \left[1 - \frac{t}{m_p^2}\right]^2\left[1 -\frac{t}{m_D^2}\right]^4 \left[1 - \frac{B t}{m_p^2}\right]^2 \left[ 1 - \frac{C t}{m_D^2}\right]^4}\, s^2   \left({-}i\,\alpha'_{\mathbb{P}} s\right)^{\epsilon_{\mathbb{P}} + \alpha'_{\mathbb{P}} t }\left(i\,\alpha'_{\mathbb{O}} s\right)^{\epsilon_{\mathbb{O}} + \alpha'_{\mathbb{O}} t },
\label{sq-amp-po-ppbar}
\end{eqnarray}
where the Regge limit, $s\ggg t,\,m_p^2$ has been applied for the approximations to obtain the final results. Here we have used the following normalizations and sum over spin of the spinors as
\begin{eqnarray}
\bar u_r(p,m_p)\,u_s(p,m_p) &=& 2\,m_p\,\delta_{rs}\,,\qquad\quad  
\sum_{r} u_r(p,m_p)\,\bar u_r(p,m_p) = p\!\!\!/ + m_p\,, 
\nonumber\\
\bar v_r(p,m_p)\,v_s(p,m_p) &=& -2\,m_p\,\delta_{rs}\,,\qquad\,\,  
\sum_{r} v_r(p,m_p)\,\bar v_r(p,m_p) = p\!\!\!/ - m_p\,, 
\end{eqnarray}
where $r$ and $s$ are spin indices of the spinors. As results, we note that $\sum_{\rm spin}\,\big|\mathcal{M}_{X}^{pp}\big|^2 = \sum_{\rm spin}\,\big|\mathcal{M}_{X}^{p\bar p}\big|^2$ with $X =\mathbb{P},\mathbb{O}$.
\\\\
Next, we will present the scalar amplitudes of the pomeron and odderon exchanges as $\mathcal{A}_{\mathbb{P}}$ and $\mathcal{A}_{\mathbb{O}}$ respectively. Having use the results in Eqs.(\ref{sq-amp-p-pp}) and (\ref{sq-amp-o-pp}), they are written by,
\begin{eqnarray}
\mathcal{A}_{\mathbb{P}}(s,t) &=& \frac{{4}\,g_{\mathbb{P}}^2 \left[1 - \frac{t}{4m_p^2}\frac{\mu_p}{\mu_N}\right]^2}{\left[1 - \frac{t}{m_p^2}\right]^2 \left[1 -\frac{t}{m_D^2}\right]^4 }\, s  \left(-i\,\alpha'_{\mathbb{P}} s\right)^{\epsilon_{\mathbb{P}} + \alpha'_{\mathbb{P}} t } ,
\label{scalar-amp-p}\\
\mathcal{A}_{\mathbb{O}}(s,t) &=& \frac{8\,g_{\mathbb{O}}^2\, \left[1- \frac{At}{4m_p^2}\frac{\mu_p}{\mu_N}\right]^2}{{3} \left[1 - \frac{B t}{m_p^2}\right]^2 \left[ 1 - \frac{C t}{m_D^2}\right]^4} \, s \left(-i\,\alpha'_{\mathbb{O}} s\right)^{\epsilon_{\mathbb{O}} + \alpha'_{\mathbb{O}} t } ,
\label{scalar-amp-o}
\end{eqnarray}
where the amplitude of the $pp$ scattering is given by $\mathcal{A}^{pp}(s,t) = \mathcal{A}_{\mathbb{P}}(s,t) - \mathcal{A}_{\mathbb{O}}(s,t) $\,.
According to the optical theorem, one can write the total cross-section formula as,
\begin{eqnarray}
\sigma_{\rm tot}^{pp} &=& \frac{1}{s}\,{\rm Im}\,\mathcal{A}^{pp}(s,0) = \frac{1}{s}\,{\rm Im}\,\Big[ \mathcal{A}_{\mathbb{P}}(s,0) - \mathcal{A}_{\mathbb{O}}(s,0)\,\Big] 
    = {\rm Im}\,\Big[ {4}\,g_{\mathbb{P}}^2\, \left(-i\,\alpha'_{\mathbb{P}} s\right)^{\epsilon_{\mathbb{P}}} 
- \frac{8}{{3}} \,g_{\mathbb{O}}^2 \left(-i\,\alpha'_{\mathbb{O}} s\right)^{\epsilon_{\mathbb{O}}} \Big]\,.
\label{tot-cross-section}
\end{eqnarray}
In the following subsection, we will use the total cross-section in Eq.(\ref{tot-cross-section}) to compare with the data from the TOTEM collaboration at the TeV scale after the model parameters are chosen.  

\subsection{Parameters fitting and discussion}
\begin{figure}
\includegraphics[width=14cm,angle=0]{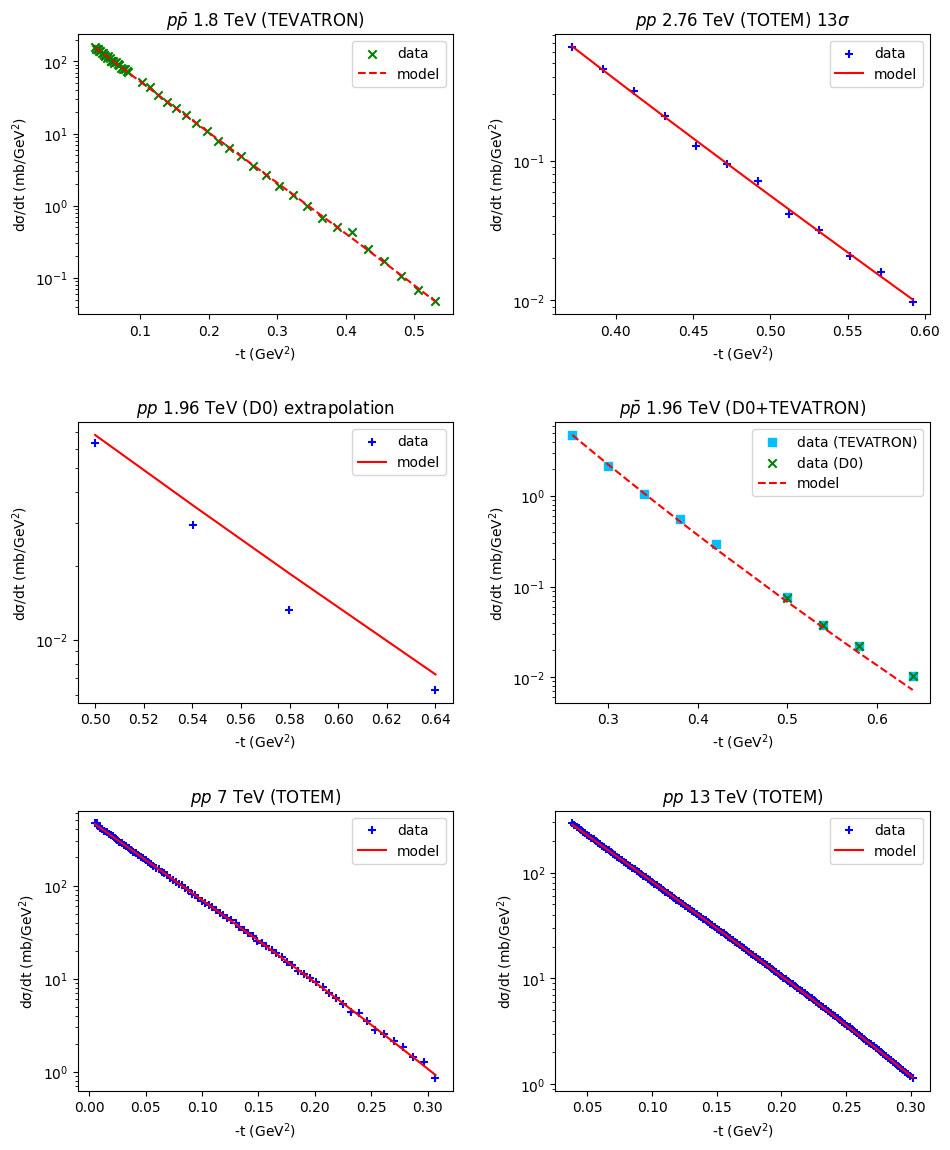}
\caption{The best fit plots of the differential cross-section of the $pp$ (blue plots) and $p\bar p$ (green plots) vs the model results with the parameters from the combined data set fit (see the last column in Table.\ref{tab:fitparam}).}
\label{diff-cross}
\end{figure}
In this section, we perform curve fitting of the model parameters with experimental data. As mentioned in the section \ref{sec-2}, we have six free parameters i.e., the odderon-$pp$ coupling constant $g_{\mathbb{O}}$, $\epsilon_{\mathbb{O}}$, $\alpha_{\mathbb{O}}'$ and the modified odderon-$pp$ form-factor parameters $A$, $B$ and $C$. The observed value for differential cross sections of $pp$ and $p\bar{p}$ scattering, ${d\sigma}^{\rm obs}/{dt}$, come from various experiments with the center of mass energy ranging from $1.8$ TeV \cite{E811:1998got} for $p\bar p$, $1.96$ TeV \cite{TOTEM:2020zzr} for $pp$ and \cite{D0:2012erd} for $p\bar p$, $2.76$ TeV \cite{TOTEM:2018psk} with $13\sigma$ and $4.3\sigma$, $7$ TeV \cite{TOTEM:2013lle} and $13$ TeV \cite{TOTEM:2017asr} for $pp$ scattering. Since we are interested in the small $-t$ limit, we only use the observed data with the linear relation between the differential cross-section and the momentum exchange because the effect of the pomeron and odderon is highly manifested in the linear regime of the differential cross-section. We define $\chi^2$ function as
\begin{equation}
    \chi^2(\alpha_i) = \sum_{j=1}^N \left(\frac{\frac{d\sigma}{dt}(\alpha_i)^{\rm model}_j - \frac{d\sigma}{dt}^{\rm obs}_j}{\frac{d\sigma}{dt}^{\rm obs}_j}\right)^2,
\end{equation}
where $\alpha_i$ are six parameters $(g_{\mathbb{O}},\,\epsilon_{\mathbb{O}},\,\alpha_{\mathbb{O}}',\,A,\,B,\,C)$ and $j = 1,\ldots, N$ is the index of the data points associated with the momentum exchange, $-t$. In order to obtain the best fit parameters, we minimize $\chi^2$ functions using \texttt{iminuit}\cite{iminuit,James:1975dr}. The results are shown in the table \ref{tab:fitparam}. Note that the errors are calculated using the Hessian matrix where more details will be provided in Appendix \ref{ch:error}. 

The central values of $g_{\mathbb{O}}$, $\alpha_{\mathbb{0}}'$, $\epsilon_{\mathbb{O}}$, $A$, $B$ and $C$ are consistent among various data sets.
The odderon mass which can be calculated from its Regge trajectory at the pole, $t=m_{\mathbb{O}}^2$ with $\alpha_{\mathbb{O}}\big( m_{\mathbb{O}}^2\big) = J = 3$ as
\begin{equation}
    m_{\mathbb{O}}(J=3) = \sqrt{\frac{J-1-\epsilon_{\mathbb{O}}}{\alpha_{\mathbb{0}}'}}\,\Bigg |_{J=3},
    \label{odd-mass}
\end{equation}
are consistent due to its dependency on $\epsilon_{\mathbb{O}}$ and $\alpha_{\mathbb{0}}'$. 

The quality of parameter fitting can be determined using the minimized $\chi^2$ per degree of freedom. Among the available data sets, the best parameters with sufficient statistics come from the $pp$-TOTEM 13 TeV data. However, using this particular data alone leads to an overfitting problem, i.e., these parameters lead to unsatisfying fits with $p\bar{p}$ data sets. We, therefore, take a more global analysis using the combined $\chi^2$ function of all available data sets. We then use the parameter fitting from the combined data set as the representation of our model. The model differential cross sections comparing with experimental data are shown in Fig \ref{diff-cross}. We have provided the statistical error analysis in detail in Appendix \ref{ch:error}. 

\begin{figure}
\includegraphics[width=9cm,height=7.5cm,angle=0]{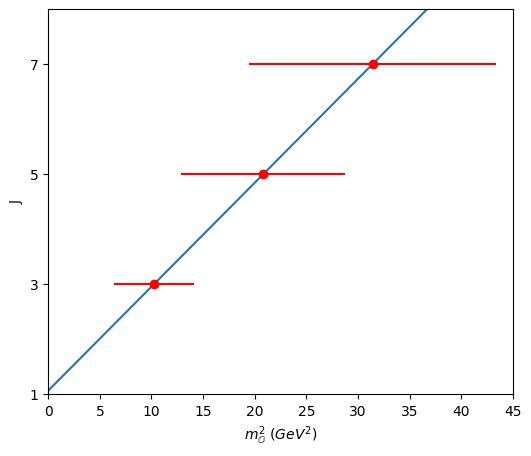}
\caption{The Chew-Frautschi plot of the odderon Regge trajectory is depicted by using the $\color{black} \alpha_{\mathbb{O}}'=0.189$ GeV$^{-2}$ and $\color{black} \epsilon_{\mathbb{O}}=0.062$ from combined data set fitting in Table\ref{tab:fitparam}. The masses of the odderon with spin-3, 5, and 7 including the error of the fitting parameter estimations with red dots are given by Eq.(\ref{odderon-masses}). }
\label{odderon-trajectory}
\end{figure}

By using the Eq.(\ref{odd-mass}) with the best-fit parameters, $\epsilon_{\mathbb{O}}$ and $\alpha_{\mathbb{0}}'$ from the combined data set, one can determine the masses of the odderons with $J = 3,\,5,\,7$ as shown below,
\begin{eqnarray}
m_{\mathbb{O}}(J^{PC}=3^{--}) &=& 3.201 \pm 0.609\;{\rm GeV}\,,
\nonumber\\ 
m_{\mathbb{O}}(J^{PC}=5^{--}) &=& 4.563 \pm 0.868\;{\rm GeV}\,,
\nonumber\\
m_{\mathbb{O}}(J^{PC}=7^{--}) &=& 5.603 \pm 1.066\;{\rm GeV}\,.
\label{odderon-masses}
\end{eqnarray}
We note that the lowest mass (pole position) of the tensor odderon with spin-3 is around 3 GeV. In addition, the Chew-Frautschi plot of the odderon Regge trajectory is depicted in Fig.\ref{odderon-trajectory}. The odderon mass results in the present work are consistent with Ref.\cite{Szanyi:2019kkn}. In that work, the authors considered the odderons as the oddballs in the double pole Regge model with spin-3, 5, and 7. Then, the masses of the odderon are extracted from the experimental data. 
{According to the literature review, we found that the theoretical estimation such as SU(3) lattice QCD for isotropic and anisotropic cases \cite{Meyer:2004jc,Chen:2005mg,Morningstar:1999rf}, Wilson loop approach \cite{Kaidalov:1999yd}, vacuum correlation method in QCD \cite{Kaidalov:1999de,Kaidalov:2005kz}, QCD sum rules \cite{Narison:1984hu,Chen:2021bck,Chen:2021cjr}, relativistic many body framework \cite{Llanes-Estrada:2005bii} give the ranges of odderon masses as $3.5-4.5$ GeV for spin-3, $5.0-5.5$ GeV for spin-5, and $6.0-6.5$ GeV for spin-7.} 
{We note that the theoretical model estimations in the literature of the odderon masses are a bit heavier than the mass estimations from the data in this work by about 0.3 GeV. However, the spin-3 odderon mass from the double pole Regge model gives $m_{\mathbb{O}}^{DP} = 3.001$ GeV \cite{Szanyi:2019kkn} which is lighter than  our work.} {From the results in Table \ref{tab:fitparam}, the odderon trajectory slope, $\color{black} \alpha_{\mathbb{O}}' = 0.189$ GeV$^{-2}$, and the pomeron slope, $\alpha_{\mathbb{P}}'$ = 0.25 GeV$^{-2}$, coming from Donnachie-Landschoff fit \cite{Donnachie:1984xq,Donnachie:1992ny} are compatible with approximation $\alpha_{\mathbb{O}}' \approx \alpha_{\mathbb{P}}'$.} 

On the other hand, we obtain $\color{black} \epsilon_{\mathbb{O}} = 0.0620\ll 1$. As a result, the best-fit value of $\alpha_{\mathbb{O}}'$ and $\epsilon_{\mathbb{O}}$ in this work correspond to the assumptions in Ref.\cite{Ewerz:2013kda} that $\alpha_{\mathbb{O}}' \approx \alpha_{\mathbb{P}}'$ and $\epsilon_{\mathbb{O}} \leq \epsilon_{\mathbb{P}}\,(=0.0808)$ which use for fixing the parameters $\alpha_{\mathbb{O}}'$ and $\epsilon_{\mathbb{O}}$ due to the lack of data used to constrain at that moment.

\begin{figure}
\includegraphics[width=12cm,height=9cm,angle=0]{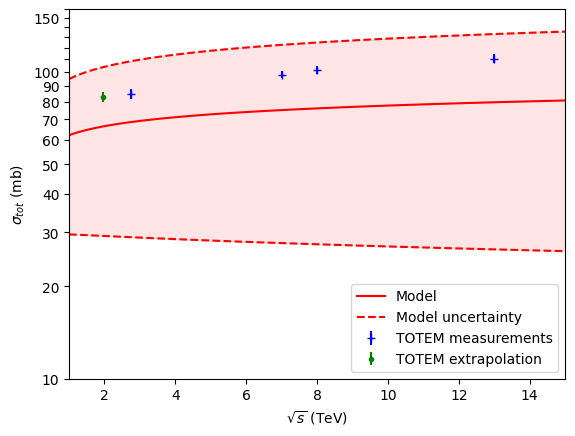}
\caption{The total cross-section of the $pp$ scattering with pomeron spin-2 and odderon spin-3 exchanges in Eq.\ref{tot-cross-section} is plotted by using the averaged fitted parameters from the combined data set in Table.\ref{tab:fitparam}. As a result, our model is compatible with the TOTEM $pp$ results at the TeV region. Moreover, the model also agrees with the extrapolation of the TOTEM data at 1.96 TeV for the $pp$ scattering. All data points are laid on the model prediction band.} 
\label{total-cross}
\end{figure}

We close this section by considering the total cross-section of the $pp$ in our model. The relevant parameters from the combined data set are substituted to the total cross-section formula in Eq.(\ref{tot-cross-section}). Then, we plot the total cross section as a function of c.m. energy ($\sqrt{s}$) as shown in Fig.\ref{total-cross} 
and we found that our model of the odderon as Regge oddball spin-3 is compatible with the TOTEM data for $pp$ scattering at the TeV regime. In particular, the extrapolation of TOTEM data for the $pp$ total cross-section at 1.96 TeV {is also laid within the error band of our model.} Using the best fit parameters $\epsilon_{\mathbb{P}} = 0.0808$ and $\color{black} \epsilon_{\mathbb{O}} = 0.0620$, in addition, the total cross-section of the present work in Eq.(\ref{tot-cross-section}) has been checked numerically and it also corresponds to a Froissart bound, i.e., $\sigma_{\rm tot}^{pp} \leq (\ln s)^2$ at $s \to \infty$ limit. 
\begin{table}[]
    \centering
    \begin{tabular}{l|c|c|c|c|c}
        \hline
         Description                         &   $\chi_{}^2$ &   $d.o.f$ & $g_{\mathbb{O}}^{}$   & $\alpha_{\mathbb{O}}'$   & $\epsilon_{\mathbb{O}}$            \\
        \hline
            1.80 TeV $p \bar{p}$ \cite{E811:1998got}                            &  0.076 &     49 & \multicolumn{1}{|r|}{9.346(±0.563)}  & 0.188(±0.040) & 0.055(±0.009) \\
            1.96 TeV $p\bar{p}$ \cite{D0:2012erd}, $pp$  \cite{TOTEM:2020zzr}   &  0.516 &     17 & \multicolumn{1}{|r|}{14.927(±1.754)} & 0.158(±0.036) & 0.064(±0.018) \\
            2.76 TeV $pp$ \cite{TOTEM:2018psk} (13$\sigma$)                     &  0.036 &     12 & \multicolumn{1}{|r|}{25.114(±3.574)} & 0.199(±0.042) & 0.070(±0.020) \\
            7.00 TeV $pp$ \cite{TOTEM:2013lle}                                  &  0.048 &     83 & \multicolumn{1}{|r|}{9.888(±0.468)}  & 0.200(±0.047) & 0.060(±0.006) \\
            13.0 TeV $pp$ \cite{TOTEM:2017asr}                                  &  0.009 &    150 & \multicolumn{1}{|r|}{9.725(±0.346)}  & 0.200(±0.024) & 0.063(±0.004) \\
        \hline
            -AVERAGES-                                         &  \multicolumn{2}{l|}{}           & 13.800(±1.112) & 0.189(±0.038) & 0.062(±0.011) \\
        \hline
        \hline
         Description                                                            & \multicolumn{2}{c|}{A}              & B              & C             & $m_{\mathbb{O}}$        \\
        \hline
            1.80 TeV $p \bar{p}$ \cite{E811:1998got}                            & \multicolumn{2}{l|}{-1.340(±0.193)} & \multicolumn{1}{|r|}{0.242(±0.262)}  & 0.436(±0.117) & 3.213(±0.611) \\
            1.96 TeV $p\bar{p}$ \cite{D0:2012erd}, $pp$  \cite{TOTEM:2020zzr}   & \multicolumn{2}{l|}{-0.855(±0.195)} & \multicolumn{1}{|r|}{1.448(±0.394)}  & 1.044(±0.151) & 3.497(±0.884) \\
            2.76 TeV $pp$ \cite{TOTEM:2018psk} (13$\sigma$)                     & \multicolumn{2}{l|}{-1.430(±0.167)} & \multicolumn{1}{|r|}{7.730(±1.374)}  & 0.064(±0.113) & 3.112(±0.773) \\
            7.00 TeV $pp$ \cite{TOTEM:2013lle}                                  & \multicolumn{2}{l|}{-2.388(±0.265)} & \multicolumn{1}{|r|}{-0.060(±0.312)} & 0.469(±0.144) & 3.116(±0.519) \\
            13.0 TeV $pp$ \cite{TOTEM:2017asr}                                  & \multicolumn{2}{l|}{-2.342(±0.145)} & \multicolumn{1}{|r|}{-0.087(±0.175)} & 0.455(±0.082) & 3.113(±0.291) \\
        \hline
             Averages                                                           & \multicolumn{2}{l|}{-1.671(±0.221)} & 1.855(±-2.121) & 0.493(±0.264) & 3.201(±0.609) \\
        \hline
        \end{tabular}
    \caption{Summary of best fit parameters for each data set.}
    \label{tab:fitparam}
\end{table}

\section{Conclusions}\label{sec-4}
In this work, we considered the odderon as the Regge oddball spin-3. The existence of odderon can be observed in a study of the difference between $pp$ and $p\bar p$ elastic scattering. We, therefore, investigate $pp$ and $p\bar p$ scattering at the Regge limit by including the pomeron and odderon exchanges in the present work. The effective Lagrangians of the processes are constructed and the standard perturbative QFT method is used to calculate the relevant observables in this work. The pomeron and odderon are identified as the Regge tensor glueballs and oddballs with spin-2 and -3, respectively. We have employed the Donnachie-Landschoff ansatz for the pomeron propagator and the Regge trajectory with the electromagnetic type of the pomeron-$pp$ form-factor. Furthermore, we also modified the electromagnetic type of the odderon-$pp$ by introducing the additional 3 free parameters. There are six free parameters of the model in this work $(g_{\mathbb{O}},\,\epsilon_{\mathbb{O}},\,\alpha_{\mathbb{O}}',\,A,\,B,\,C)$. Having performed a careful statistical analysis, all free parameters have been fixed by fitting with all combined data of the $pp$ and $p\bar p$ differential cross-sections at the TeV regime see results in Table \ref{tab:fitparam}. After fixing the free parameters in the present work, the masses of the odderon spin-3 and their excited states for spin-5 and -7 are estimated from its Regge trajectory by using the best fit of the combined data set. Considering the best-fit results in Table \ref{tab:fitparam}, the oddereon Regge trajectory parameters are found to be $\color{black} \alpha_{\mathbb{O}}'=0.189\,\pm\,0.038$ and $\color{black} \epsilon_{\mathbb{O}} = 0.062\,\pm\,0.011$. These results are compatible with the assumptions in Ref.\cite{Ewerz:2013kda} that used to estimate those two parameters as $\color{black} \alpha_{\mathbb{O}}'\approx \alpha_{\mathbb{P}}'$ and $\epsilon_{\mathbb{O}} \leq \epsilon_{\mathbb{P}}$. As a result, we found that the tensor odderon masses are heavier than the phenomenological approach by using the double pole Regge model extracted from the experimental data \cite{Szanyi:2019kkn}. On the other hand, the odderon masses in this work are lighter than other model theoretical calculations in the literature for all odderons along their trajectory by about 0.3 GeV. Having used the best-fit parameters, the total cross-sections also agree with the TOTEM data in the TeV regime and its extrapolation from D0 of the $pp$ scattering at 1.96 TeV. The odderon spin-3 contribution also provided the amplitude in Regge limit that satisfied the Froissart bound. According to our findings in this work, we can conclude that the odderon could be the tensor Regge oddball spin-3 particle. Further studies to confirm our conclusion are needed to investigate other scattering processes such as polarized proton-proton scattering, meson-baryon scattering, and photoproduction. We plan to do this in the forthcoming future work.


\acknowledgments
DS is supported by the Fundamental Fund of Khon Kaen University and DS has received funding support from the National Science, Research and Innovation Fund. The Mindanao State University - Iligan Institute of Technology is also acknowledged through its research and extension support extended to J.B. Magallanes for his travel to Khon Kaen University, Thailand. PS, CP, and DS are financially supported by the National Astronomical Research Institute of Thailand (NARIT). CP and DS are supported by Thailand NSRF via PMU-B [grant number B05F650021]. CP is also supported by Fundamental Fund 2565 of Khon Kaen University and Research Grant for New Scholar, Office of the Permanent Secretary, Ministry of Higher Education, Science, Research and Innovation under contract no. RGNS
64-043.

The authors acknowledge the National Science and Technology Development Agency, National e-Science Infrastructure Consortium, Chulalongkorn University and the Chulalongkorn Academic Advancement into Its 2nd Century Project, NSRF via the Program Management Unit for Human Resources \& Institutional Development, Research and Innovation [grant numbers B05F650021, B37G660013] (Thailand) for providing computing infrastructure that has contributed to the research results reported within this paper. URL:www.e-science.in.th.


\bibliography{references.bib}

\appendix
\section{Error analysis} \label{ch:error}
\begin{figure}
\includegraphics[width=15cm,angle=0]{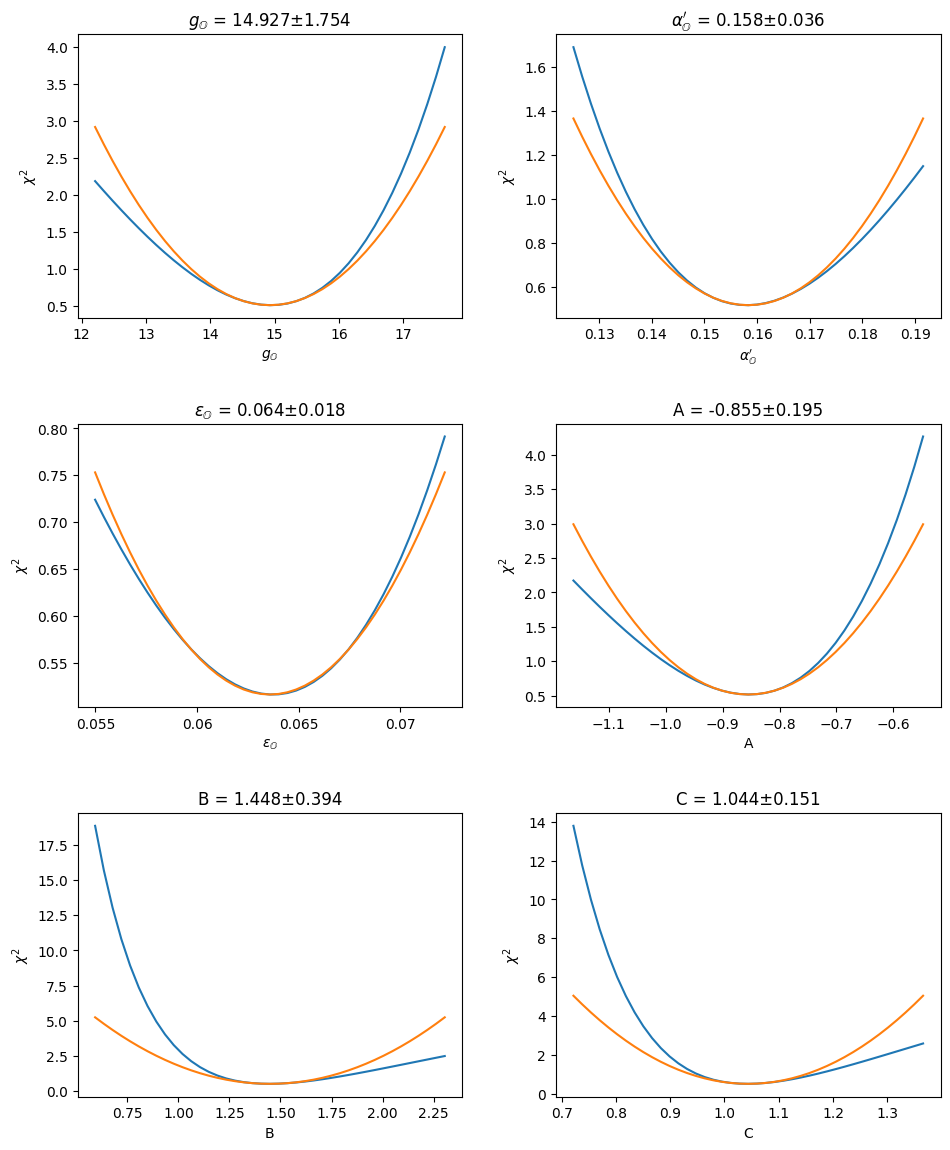}
\caption{The comparison plots between the parabolic function using $\chi^2 = \chi^{\rm min} + \frac{\left(\alpha_i - \alpha_i^{\rm min}\right)^2}{\sigma^2}$ (orange line) and the real $\chi^2$ function (blue line). The combined data set \textcolor{black}{of 1.96 GeV ($pp,p\bar{p}$)} is used to obtain the parameter fit in these plots. Each panel represents the variation of the $\chi^2$ function in each particular direction of the parameter space.}
\label{chi2}
\end{figure}

We provide the detail of the error estimation method in this section. The errors for parameter fitting in this analysis are handling outside the \texttt{iminuit} package due to inaccuracy of the results. The errors shown in Table \ref{tab:fitparam} are therefore improved by the following. Consider the Taylor expansion around the minimum of the $\chi^2$ function,
\begin{equation}
    \chi^2(\alpha_{i}) \approx \chi^2(\alpha_{i}^{\rm min}) + \frac{1}{2}\left(\alpha_i - \alpha_{i}^{\rm min}\right)^2 \frac{\partial \chi^2}{\partial \alpha_i}|_{\alpha_i = \alpha_{i}^{\rm min}} + \mathcal{O}\left(\left(\alpha_i - \alpha_{i}^{\rm min}\right)^3\right).
\end{equation}
We can approximate the error of parameter estimation, $\sigma_i$, using the width of parabolic function defined as
\begin{equation}
    \frac{1}{\sigma^2_i} = \frac{1}{2} \frac{\partial \chi^2}{\partial \alpha_i}|_{\alpha_i = \alpha_{i}^{\rm min}},
\end{equation}
which is also the diagonal component of the Hessian matrix. The second derivative of the $\chi^2$ function is obtained via the finite difference method
\begin{equation}
    \frac{\partial \chi^2}{\partial \alpha_i}|_{\alpha_i = \alpha_{i}^{\rm min}} \approx \frac{\chi^2(a_i^{\rm min}+\Delta a_i)+\chi^2(a_i^{\rm min}-\Delta a_i) - 2\chi^2(a_i^{\rm min})}{\Delta a_i^2},
\end{equation}
where $\Delta a_i$ is chosen to be sufficiently small compared to the value of the error.
The validity of the approximation is then confirmed by the comparison between the parabolic functions and the real $\chi^2$ function shown in Fig.\ref{chi2}. One can see that both functions agree very well within the range of error estimations.

\end{document}